\documentclass[%
twocolumn,
 aip,
 jmp,%
 amsmath,amssymb, 
 reprint,%
]{revtex4-1}

\usepackage{graphicx}
\usepackage{dcolumn}
\usepackage{bm}
\usepackage[usenames,dvipsnames]{color}
\usepackage{multirow}
\usepackage[hyperfootnotes=false]{hyperref}
\usepackage[hyperfootnotes=false]{hyperref}
\usepackage{physics}
\hypersetup{
 colorlinks,
 citecolor=Violet,
 linkcolor=Red,
 filecolor= green,
 urlcolor=Blue}

\begin{document}


\title{Strain induced large enhancement of thermoelectric figure-of-merit ($ZT\sim 2$) in transition metal dichalcogenide monolayers ZrX$_2$ (X= S, Se, Te) }

\author{Ransell D'Souza}%
 \email {ransell.d@gmail.com (Present address: Tyndall National Institute, University College Cork, Cork, Ireland)}
\altaffiliation{ }
\author{Sugata Mukherjee}%
 \email{sugata@bose.res.in; sugatamukh@gmail.com (Corresponding Author)}
\affiliation{ 
Department of Condensed Matter Physics \& Materials Science, S.N. Bose National Centre for Basic Sciences, Block JD, Sector III,
Salt Lake, Kolkata 700098, India}

\author{Sohail Ahmad}%
 \email{sohailphysics@yahoo.co.in}
\affiliation{%
Department of Physics, Faculty of Science, King Khalid University, Abha, Saudi Arabia
}%

\date{\today}

\begin{abstract}
Two-dimensional group {\rm IV} transition-metal dichalcogenides have  encouraging thermoelectric applications since their electronic and lattice properties can be manipulated with strain. In this paper, we report the thermoelectric parameters such as electrical conductivity, Seebeck coefficients, electrical relaxation times, and {\color{black}the mode dependent contributions to the }lattice thermal conductivity of ZrX$_2$ (X = S, Se, Te) from {\color{black}first principles methods.}
Our calculations indicate that due to tensile strain, the powerfactor increases while simultaneously decreasing the lattice thermal conductivity thus enhancing the thermoelectric figure of merit. Tensile strain widens the bandgap which corresponds to higher powerfactor. The lattice thermal conductivity decreases due to the stiffening of the out-of-plane phonon modes thus reducing the anharmonic scattering lifetimes and increasing the thermoelectric figure-of-merit.
%
\end{abstract}

\maketitle

\section{INTRODUCTION}
Thermoelectric (TE) conversion can be a very simple and feasible solution to the ever growing energy crisis, since the waste energy from the by-products of solid state devices can be directly converted into electricity.
TE materials, therefore, have encouraging applications in power generations and solid-state cooling.
Due to different scattering mechanisms in phonons at lower dimensions, low-dimensional materials have been preferred over their corresponding bulk counterparts for an enhanced TE performance \cite{he13,zhang14,park16}.
For example, monolayers of MX$_2$ (M=Mo,W; X=S,Se) demonstrate better thermoelectric performances than their corresponding bulks \cite{buscema13,huang13,wickramaratne14,babaei14}.
With the advancement in nano-technology, two-dimensional (2D) materials have been produced successfully \cite{coleman11,jeong15}.
Two dimensional ZrS$_2$ and ZrSe$_2$ have been experimentally synthesized by Zeng {\it et al.}\cite{zeng11} and Sargar {\it et al.} \cite{sargar09}, respectively. Very recently, Tsipas {\it et al.} \cite{tsipas18} have grown 2D ZrTe$_2$ using molecular beam epitaxy on InAs(111)/Si(111) substrates. Development in the synthesis of these low dimensional materials and the fact that their low dimensional properties have improved thermoelectric properties have opened up new opportunities in group {\rm IV} transition metal dichalcogenide (TMD) in the field of thermoelectrics.

The conversion efficiency of any thermoelectric material is characterized by its figure-of-merit,
$ZT = \frac{S^2\sigma}{\kappa}T$, a dimensionless parameter that depends on the voltage induced by temperature
gradient known as Seebeck coefficient ($S$), electrical conductivity ($\sigma$), absolute temperature ($T$) and total thermal conductivity ($\kappa$) having contributions from both electrons and lattice vibrations (phonons). Enhancement of $ZT$ is a challenging endeavor since $\sigma$, $S$ and $\kappa$ are unfavorably related to each other {\it i.e.} larger $\sigma$ would generally lead to lower $S$ and larger $\kappa$, thus lowering $ZT$.

{\color{black}In recent years, numerous techniques to improve $ZT$ for two dimensional materials have been examined.
These techniques involve defect engineering, $n$ and $p$-type doping, and by inducing strain.}
For example, Anno {\it et al.} \cite{anno17} have enhanced the figure of merit of graphene through defect engineering. Similarly, Tabarraei {\it et al.} \cite{tabarraei15} calculated the thermal conductivity of hexagonal boron nitride with defects such as boron mono-vacancy, nitrogen mono-vacancy and Stone-Wales defects. Fei {\it et al.}\cite{fei14,fei} have reported a $ZT \sim 1$  at room-temperature (RT) for 2D-Phosphorene.
In the case of two-dimensional boron nitride, we have shown reduction in the lattice thermal conductivity by stacking \cite{RSBN17}.
We have also recently calculated the enhancement of $ZT$ of graphene by doping it with dimers of boron nitride which simultaneously increases the power factor ($S^2\sigma$) \cite{RDSM17} and decreases the lattice thermal conductivity \cite{RDSM18}.

There are various methods for doping in 2D materials. 
For example, $n$-type ($p$-type) doping can be achieved by donating (extracting) electrons from the 2D material \cite{mouri13,kiriya14}. $n$-type ($p$-type) doping can also be achieved by replacing the site atom with another atom having more (less) valence electrons than that of the site atoms \cite{laskar14}. These techniques have been done for MoS$_2$ and hence can be easily carried out on ZrX$_2$. Doping can also be achieved without randomly distributed impurities by electrolyte gating \cite{das08, ye12}.

Strain can be introduced in 2D materials by lattice mismatch between the substrates and epitaxial thin films \cite{gao07}. By transferring a 2D material onto a substrate, strain can be induced on the 2D material by bending \cite{park08} or stretching \cite{kim09} the substrate.

Several 2D materials under the influence of strain demonstrate a functional technique to tune the electronic bandstructure \cite{fei14,huang13,ghosh15,srs19} and phonon dispersion \cite{hu13,zhu13,srs19}.
Inducing strain on phosphorene has been shown to considerably enhance the Seebeck coefficient \cite{zhu13}.
Solution to the phonon transport Boltzmann equation from molecular and lattice dynamics were carried out for graphene and boron-nitride to study their non-monotonic behavior of lattice thermal conductivity \cite{taishan15}.
While the lattice thermal conductivity of graphene does not vary much with small strain \cite{lindsay14, taishan15} there have been various reports on the investigation on the mode contributions to the lattice thermal conductivity \cite{lindsay14,kuang16,lindsay2011,lindsay2010}. Therefore there is a need to investigate the mode dependence contribution to the lattice thermal conductivity which, to the best of our knowledge, does not exist in literature. In this paper, we therefore calculate the contribution to the lattice thermal conductivity of ZrX$_2$ ; X = S, Se, Te. 
Moreover, we calculate the figure of merit by calculating the electron relaxation time and electrical thermoelectric parameters, the electrical conductivity and Seebeck coefficients.

Within the framework of density functional theory (DFT) and density functional perturbation theory (DFPT), we have recently reported results of the structural parameters, electronic bandstructure, and phonon dispersion for unstrained and strained (tensile and compressive) monolayers of ZrX$_2$ (X=S,Se,Te) \cite{srs19}, {\color{black} where the structure of the monolayers with lattice parameters are also given}. Using these electronic and phonon bandstructure data, electrical and lattice transport parameters of ZrX$_2$ (X=S,Se,Te) monolayers are obtained using semiclassical electrical and phonon Boltzmann transport equations (BTE), respectively. To calculate the electrical conductivity and Seebeck coefficients, the electrical BTE are applied to the band electrons derived from DFT, in the relaxation time approximation (RTA), as implemented in the BoltzTraP code \cite{boltztrap}. Similarly, the lattice thermal conductivity ($\kappa_L$) is calculated by solving the phonon BTE, applied beyond the RTA, using a real space supercell approach using the phonon dispersion derived from DFPT as implemented in the ShengBTE code \cite{ShengBTE}.
\subsection{Electrical Boltzmann transport equations}
The electrical BTE calculates the transport properties along the two orthogonal principal axes in the $xy-$plane. The $S$ and $\sigma$ values are averaged over these principal directions. The velocities of the electrons can be determined from the energies ($\varepsilon$) and wave-vector (${\bf k}$) points derived from DFT. The expression for the wave-dependent velocity ($v_{\alpha}(i,\textbf{k})$) in the $i^{th}$ band is given by,
\begin{eqnarray}
v_{\alpha}(i,\textbf{k})=\frac{1}{\hbar}\frac{\partial \varepsilon_{i,k}}{\partial k_{\alpha}},
\end{eqnarray}
where $\alpha$ denotes the velocity's component.
In the RTA, $\sigma$ and $S$ tensors are expressed in terms of the electron velocities as \cite{boltztrap},
\begin{eqnarray}\label{sig}
{\sigma_{\alpha\beta}(T,\mu) } &=& {1 \over V} \int \tau_e e^2 v_{\alpha}(i,{\bf k})\, v_{\beta}(i,{\bf k}) [{-\partial f_\mu(T,\epsilon) \over \partial \epsilon}] d\epsilon
\end{eqnarray}
and
\begin{eqnarray}\label{S}
S_{\alpha \beta}= {\frac{1}{eT}}{{\int \tau_e v_{\alpha}(i,\textbf{k})v_{\beta}(i,\textbf{k})(\epsilon-\mu) [{-\partial f_\mu(T,\epsilon) \over \partial \epsilon}] d\epsilon}\over {\int \tau_e v_{\alpha}(i,\textbf{k})v_{\beta}(i,\textbf{k}) [{-\partial f_\mu(T,\epsilon) \over \partial \epsilon}] d\epsilon}}
\end{eqnarray}
Here, the electron relaxation time, Fermi-Dirac distribution, chemical potential, volume of the unit cell, absolute temperature and the Boltzmann constant are denoted by $\tau_e$, $f_{\mu} = \frac{1}{1+ e^{\frac{\varepsilon-\mu}{k_BT}}}$, $\mu$, $V$, $T$, and $k_B$, respectively.

\subsection{Electrical relaxation time}\label{ert}
In order to estimate the electrical relaxation time, we employ the model \cite{zahedifar18,durczewski2000} as implemented in the BoltzTrap code, where
\begin{eqnarray}\label{tau}
\tau    (E,T) = \tau_0 \bigg(\frac{E-E_{VBM/CBM}}{k_BT}\bigg)^{r - \frac{1}{2}}\bigg(\frac{T_0}{T}\bigg)^{l}.
\end{eqnarray}
Here, $E_{VBM(CBM)}$  are the valence band maximum (conduction band minimum) for the $p$-type ($n-$type) conduction.
{\color{black}$\tau_0$ is the reference lifetime, specified at some reference temperature $T_0$.}
The exponent of energy $r$, known as the scattering parameter and the exponent of the temperature dependence, $l$, have following values for different scattering mechanisms: (i) Scattering due to acoustic phonons in 2D (3D) materials, $r=\frac{3}{2}$ $(r = 2)$ and $l=0$ and (ii) $r=1$ and $l = 1$ for scattering by optical phonons at high temperatures. Moreover, a $l$ value of -3 (-4) would implement effects due to scattering based on phonon isotopes (normal phonon).
The scattering parameter $r$ can not be negative. However, if the Fermi energy is so low that scattering angles upto $\pi$ are allowed, $r = 0$.
More detailed information about the method have been reported by Durczewski and Ausloos \cite{durczewski2000}, Okuda {\it et  al.} \cite{okuda01} and Palmer {\it et  al.} \cite{palmer97}.

\subsection{Phonon Boltzmann transport equations}
The calculations of the lattice thermal conductivity ($\kappa_L$) require the second-order (harmonic) and third-order (anharmonic) inter-atomic force constants (IFCs).
The harmonic IFCs have been calculated by us earlier to obtain the phonon dispersion \cite{srs19}. We therefore have to calculate only the third-order IFCs and thereafter obtain the lattice thermal conductivity.

Expanding the potential energy ($E$) around its equilibrium energy ($E_0$), we obtain the $n^{th}$ order IFCs ($\phi$) which are the coefficients of the $n^{th}$ order term,
\begin{eqnarray}
E = E_0 + \frac{1}{2}\sum_{\substack{i j\\ \alpha \beta}}\phi_{ij}^{\alpha \beta}r_i^{\alpha}r_j^{\beta} + \frac{1}{3!}\sum_{\substack{i j k\\ \alpha \beta \gamma}}\phi_{ijk}^{\alpha \beta \gamma}r_i^{\alpha}r_j^{\beta}r_k^{\gamma} + \cdots.
\end{eqnarray}
The third order IFCs can therefore be expressed as,
\begin{eqnarray}\label{3ifc}
\phi_{ijk}^{\alpha \beta \gamma} = \frac{\partial^3 E}{\partial r^{\alpha}_i \partial r^{\beta}_j \partial r^{\gamma}_k}.
\end{eqnarray}
Using the finite difference method, equation \ref{3ifc} can be approximately expressed as,
\begin{eqnarray}
&\phi_{ijk}^{\alpha \beta \gamma}& \approx  \frac{1}{4h^2}[-F_k^{\gamma}(r_i^{\alpha}=h, r_j^{\beta}=h)+F_k^{\gamma}(r_i^{\alpha}=h, r_j^{\beta}= -h)\nonumber \\ 
&+& F_k^{\gamma}(r_i^{\alpha}=-h, r_j^{\beta}=h)-F_k^{\gamma}(r_i^{\alpha}=-h, r_j^{\beta}=-h)],
\end{eqnarray}
where, $h$ is an arbitrary small displacement from the system's equilibrium position. $F_k^{\gamma}$ is the $\gamma$ component of the force experienced by the $k^{th}$ atom.
We use the ``thirdorder.py" python script which is part of the ShengBTE package to generate different configurations having displaced atoms.
The number of configurations created depends on the number of nearest neighbor interactions, symmetry of the system, size of the supercell mesh, and size of the unit cell.
In all of our calculations, the super-cell mesh size was fixed at $3\times 3\times 3$. The number of nearest neighbor interactions was chosen to be four. 300 and 372 configurations were generated for each system with and without strain, respectively. {\color{black} Note that the difference in the number of configurations is due to the change in symmetry of the system under the influence of strain.}

After obtaining harmonic and anharmonic terms, the lattice thermal conductivity tensor is then calculated beyond the RTA using the following expression,
\begin{eqnarray}\label{kl}
\kappa_L^{\alpha \beta}=\frac{1}{k_BT^2 V N}\sum_{\lambda}f_0(f_0+1)(\hbar \omega_\lambda)^2v_{\lambda}^{\alpha} \tau_{\lambda}^0 (v_\lambda^{\beta}+\Delta_\lambda^{\beta}).
\end{eqnarray}
Here, $N$ is the total number of sampling wave vector (${\bf q}$) points (we use a different notation here because the points correspond to that of phonons).
The size of the grid planes mesh along each axis in reciprocal space was chosen to be $24 \times 24 \times 1$ for all our calculations.
 $\tau^0$ represents the relaxation time, $\lambda$ corresponds to the acoustic and optical modes and $v$ the phonon group velocity. The Bose-Einstein distribution is $f_{0} = \frac{1}{e^{\frac{\hbar \omega}{k_BT}}-1}$ where $\hbar$ and $\omega$ are the Planck's constant, and phonon frequency.
 It is evident from equation \ref{kl} that by setting $\Delta =0$ is identical to solving the phonon BTE in the RTA. Therefore, $\Delta$ is a measure how much the associated heat current of a particular phonon mode deviates from the RTA solution of the phonon BTE. $\Delta$ is a function of the three-phonon processes and the scattering probabilities from isotropic disorders. The three-phonon scattering amplitudes, which are a function of the third-order anharmonic term, is then computed from the derivatives of energy obtained from DFT using the QUANTUM ESPRESSO code \cite{giannozzi09}. Using the ShengBTE code, the relaxation times are obtained by an iterative method with the initial run as the RTA, {\it i.e.}, $\Delta =0$. Iterations stop when two successive runs for $\Delta$(and hence $\tau$) have a difference of 10$^{-5}$ Wm$^{-1}$K$^{-1}$.

In order to get an estimate of the lattice thermal conductivity as function of its sample length, we calculate the cumulative $\kappa_L$ since the mean-free path (MFP) cannot be larger in size than the physical sample length of the sheet. The cumulative $\kappa_L$ is obtained by permitting only phonons with a MFP below some certain threshold value.  The cumulative $\kappa_L$ is then fitted to the form \cite{ShengBTE},
\begin{eqnarray} \label{cum-k}
\kappa_L(L) = \frac{\kappa_{L_{max}}}{1+\frac{L_0}{L}},
\end{eqnarray}
where, $\kappa_{L_{max}}$ is the thermodynamic limit of lattice thermal conductivity, {\it i.e.}, $\kappa_L$ as $L \rightarrow \infty$. $L_0$ is a fitting parameter and $L$ is the MFP (or sample length of the sheet). {\color{black}Detailed information} and workflow has been reported by Li {\it et. al.} \cite{ShengBTE}.

\section{RESULTS AND DISCUSSIONS}
\subsection{Electrical Boltzmann Transport}
Using the Boltzmann transport equation (BTE) for the band electrons, we calculate the scaled electrical  conductivity ($\frac{\sigma}{\tau_e}$; $\tau_e$ is the electrical relaxation time), Seebeck coefficient and the powerfactor ($S^2\sigma$). The Seebeck coefficient, from its definition (Eq. \ref{S}), can be shown to be independent of its electrical relaxation time.
\begin{figure}[h!]
\centering
\includegraphics[scale=0.37]{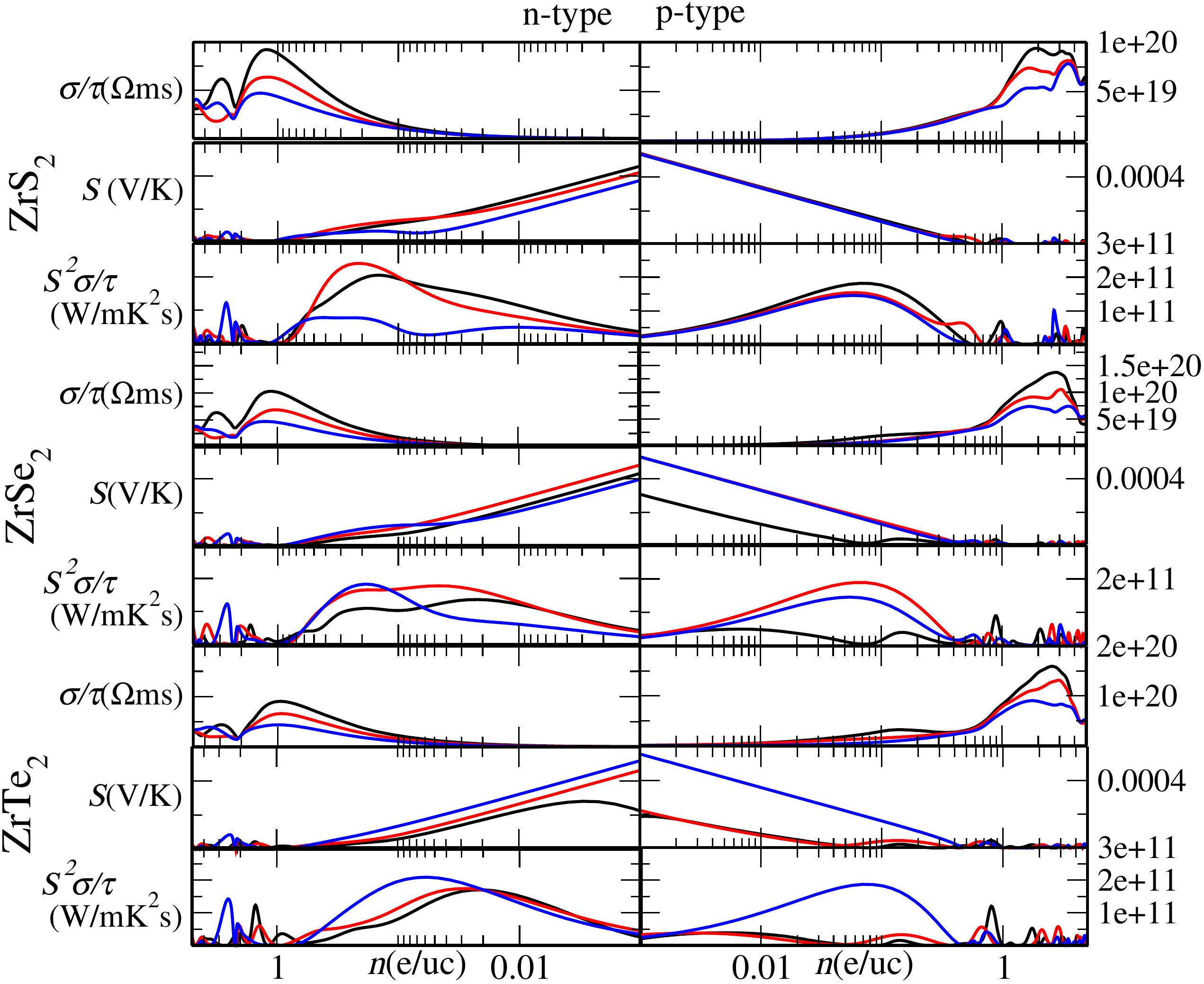}
\caption{Electronic transport parameters as a function of its charge carrier (number of electron per unit cell) for ZrX$_2$; X=S,Se,Te. The first three rows are the scaled electrical conductivity, Seebeck coefficient and the powerfactor for ZrS$_2$, respectively. Similarly, the next three are for ZrSe$_2$ and the last three are for ZrTe$_2$. The black curves represent unstrained ZrX$_2$ while red and blue curves represent a biaxial strain of 6\% and 10\%, respectively. The two coloumns correspond to $n$-type (left) and $p$-type (right) doping.}\label{pf}
\end{figure}

In Fig. \ref{pf} we present the electronic transport parameters of ZrX$_2$ as a function of its charge carriers for unstrained (black curves), 6\% biaxial strain (red curves) and 10\% biaxial strain (blue curves). There is a general trend that all $n$-type doped systems have larger powerfactor in comparison to their corresponding $p$-type doping. This behavior can be attributed to the larger effective mass in $n$-type doped systems in ZrX$_2$. Therefore, our calculations suggest that $n$-type doping would yield better results for an enhanced thermoelectric performance.

The most interesting feature is that the maximum of the powerfactor for ZrS$_2$ and ZrSe$_2$ is for a tensile strain of 6\% while that of ZrTe$_2$ is 10\%. The reason for this can be understood with the analogy of doping boron-nitride with graphene. Doping boron-nitride with graphene modulates the electronic band-gap \cite{rdsm15,rst17,grossman12} just like applying strain in ZrX$_2$. We have shown recently \cite{srs19} that the band-gap of ZrX$_2$ has a maximum at 6\% for X=S,Se while the band-gap keeps increasing with tensile strain for X=Te. Since ZrX$_2$ are either metals or semiconductors, it can be shown that a Sommerfeld expansion around the Fermi energy yields the Mott's formula, $S\propto\frac{1}{\sigma}\frac{d\sigma}{d\mu}$. Therefore, from the Mott's formula, it is seen that the {\color{black}two competing} contributions to $S$ are $\sigma^{-1}$ and $\frac{d\sigma}{d\mu}$. If the shape of the band-structure does not change much, the major contribution to $S$ would be from $\sigma^{-1}$. The bandstructure calculations does show that the shape of the bands are not much affected. However, the band gap increases with tensile strain. {\color{black}Large band gap implies a lower electrical conductivity, thus producing a larger $S$. This explains the maximum powerfactor in ZrX$_2$ for each case.}  Similar results for ZrS$_2$ \cite{lv16} and ZrSe$_2$ \cite{ding16} have been reported and are in excellent agreement with our calculations. However, they do not offer an explanation for such behavior.
We would like to stress the fact that, as against the popular belief that lower electrical conductivity is detrimental to $ZT$, Mott's formula shows that, lower conductivity for metals and semiconductors would actually increase the Seebeck coefficients thus increasing the powerfactor which is proportional to the square of S.

\begin{figure}[h!]
\centering
\includegraphics[scale=0.33]{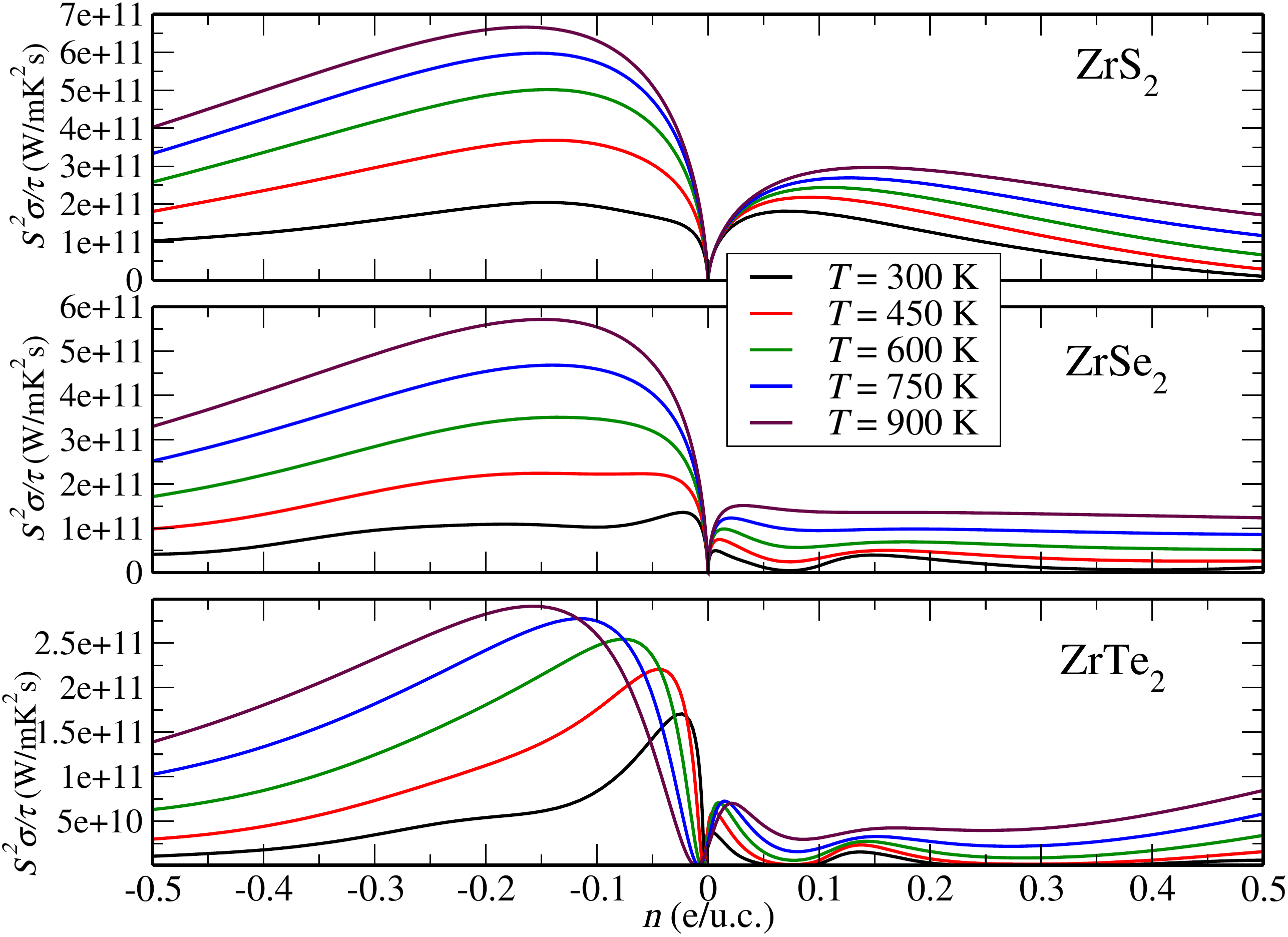}
\caption{Powerfactor of ZrX$_2$ as a function of the charge carrier at five different temperatures, $T$=300K, 450K, 600K, 750K, and 900K. }\label{pf-temp}
\end{figure}

In Fig. \ref{pf-temp}, we plot the powerfactor as a function of its charge carrier concentration at temperatures, $T=$ 300K, 450K, 600K, 750K, and 900K. A common trend in the behavior of ZrX$_2$ powerfactor is that it increases with the temperature. Moreover, only near $|n|\sim 0$,  higher the temperature, the peak of the powerfactor shifts to a higher value of $|n|$. This behavior can be ascribed to {\color{black}the more thermally excited electrons} since $|n|\sim 0$ implies that the chemical potential is near the Fermi energy. Therefore, the concentration of electrons would increase due to the availability of energy states above the ground state. Higher temperature would also {\color{black}imply more thermally excited electrons.}

We next consider the effect of tensile strains on lattice thermal conductivity of ZrX$_2$ since Seebeck coefficients and powerfactor increase due to strain thus improving the thermoelectric figure of merit.

\subsection{Effective mass and Electron Relaxation time}
In the previous sections, we presented the scaled electrical conductivity and Seebeck coefficients. However, in order to calculate the powerfactor, one would require the electrical relaxation time. Previously, we compared our scaled electrical conductivity with experiments to get an estimate for the electrical relaxation time, for example, with graphene \cite{RDSM18} and graphene/boron nitride heterostructures \cite{RDSM16}. In the case of ZrX$_2$, to the best of our knowledge, no experiments have been reported to compare our scaled electrical conductivity. Therefore, in this section, we calculate the relaxation time using equation \ref{tau}.

One would require the reference lifetime, the scattering parameter and the temperature dependence to have an accurate estimate of the electrical relaxation time. The reference lifetime can be estimated from the fact that the electron scattering rate given by the Fermi's golden rule \cite{sakurai},
\begin{eqnarray}\label{fgr}
\tau_e = \frac{\hbar}{2\pi g(E) |\bra{f}H'\ket{i}|^2},
\end{eqnarray}
where, $g(E)$ is the energy density of state and is proportional to the directional effective mass. $\bra{f}H'\ket{i}$ is the matrix element of the perturbation Hamiltonian $H'$ between final ($\bra{f}$) state and initial state ($\ket{i}$).
 Thermoelectric materials exhibit directional behavior mainly from the directional effective masses. The directional effective mass was calculated by approximating the electronic bandstructure around the valence band maximum and conduction band minimum to be parabolic,
\begin{eqnarray}
E = \frac{\hbar^2}{2}\bigg(\frac{k_x^2}{m_x} + \frac{k_y^2}{m_y}\bigg).
\end{eqnarray}
Here, $m_x$ and $m_y$ are the principal directional effective masses. $k_x$ and $k_y$ are the components of the wavevector with a magnitude expressed as, $k = \sqrt{k_x^2 + k_y^2}$. We define a new wavevector, $k'$, and effective mass, $m'$ for the system by the relation,
\begin{eqnarray}\label{E}
E = \frac{\hbar^2 k'^2}{2m^{'}} = \frac{\hbar^2 \big(k_x'^2 + k_y'^2 \big)}{2m^{'}}
\end{eqnarray}

The relation between the new proposed wavevector and effective mass with the components of wavevector and directional effective masses is thus given by,
\begin{eqnarray}
k_x(y) = \sqrt{\frac{m_{x(y)}}{m'}}k_{x(y)}'.
\end{eqnarray}
The infinitesimal surface area element for a two-dimensional material is expressed,
\begin{eqnarray}
dk &=& dk_xdk_y = \sqrt{\frac{m_x m_y}{m'^2}}dk_x'dk_y' \nonumber \\ &=& \sqrt{\frac{m_x m_y}{m'^2}}2\pi k'dk'.
\end{eqnarray}

The density of states is then calculated by counting the number of states between $k$ and $k + dk$ in two-dimensional space divided by the smallest area of the wavevector in a crystal, $\big(\frac{2\pi}{L}\big)^2$.
Accounting for the electron spin, Pauli Exclusion Principle, the density of states, $g(k) dk$, is expressed as,
\begin{eqnarray}\label{dos}
g(k)dk = \bigg(\frac{\sqrt{m_xm_y}}{\pi}\bigg)\frac{k'dk'}{m'}.
\end{eqnarray}
By differentiating Eq. \ref{E} and by substituting $\frac{k'dk'}{m'}$ in Eq. \ref{dos}, the energy density of states is expressed as,
\begin{eqnarray}\label{Edos}
g(E) = \bigg(\frac{\sqrt{m_xm_y}}{\pi}\bigg) \bigg(\frac{1}{\hbar^2}\bigg)
\end{eqnarray}
From Eq. \ref{fgr}, the scattering lifetime is given by,
\begin{eqnarray}\label{taue}
\tau_e = \Big(\frac{\pi}{\sqrt{m_xm_y}}\Big)\Big(\frac{\hbar}{2\pi}\Big)\Big( \frac{\hbar^2}{|\bra{f}H'\ket{i}|^2} \Big)
\end{eqnarray}
The calculation of the matrix element in Eq. \ref{taue} is beyond the scope of this paper. However, it is a constant and its value would be governed by the deformation potential.
 The deformation potential have been shown to be $\sim 1$ eV for ZrX$_2$ materials. For example the deformation potentials for ZrS$_2$ and ZrSe$_2$ are 1.52 eV and 1.25 eV, respectively \cite{huang16}.
We thus choose Eq. \ref{taue} as the reference lifetime in our calculations with $|\bra{f}H'\ket{i}|^2 \sim 1$ eV.
\begin{figure}[h!]
\includegraphics[scale=0.35]{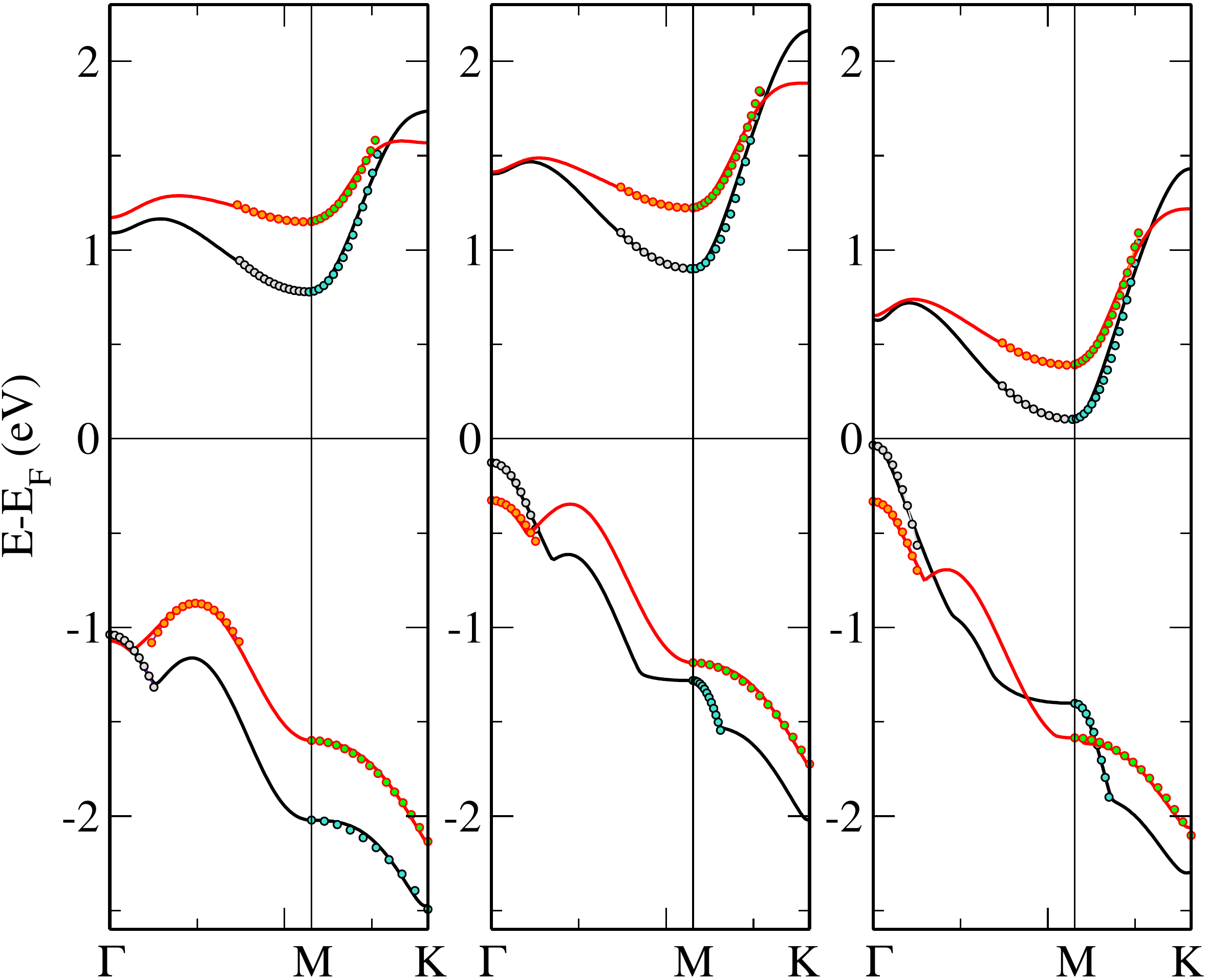}
\caption{The valence band maximmum and conduction band minimum for unstrained and 6\% strain ZrX$_2$. The circular points refer to that of the parabolic curve (Eq. \ref{E}).}\label{eff_mass}
\end{figure}

In Fig. \ref{eff_mass}, we plot the band structure for the valence band maximum and conduction band minimum for unstrained and 6\% strain ZrX$_2$. The black (red) curves refer to that of the unstrained (strained) systems.
The circular points refer to that of the parabolic curve as shown in Eq. \ref{E} with $m'$ as the fitting parameter. The values of $m'$ are shown in Table \ref{tab}. The parabolic fit represent the actual band structure well for values around the valence band maximum and conduction band minimum implying that the effective mass at valence band maximum and conduction band minimum are extremely accurate.

\begin{table}[h]
\caption{\label{tab} Effective masses of unstrained ZrX$_2$ and under 6\% tensile strain at the valence band maximum and conduction band minimum for two perpendicular directions along the high symmetric points, $\Gamma$M and KM.}
\resizebox{0.4\textwidth}{!}{
\begin{tabular}{cc|cc|cc}
\cline{3-6}
& & \multicolumn{2}{ c| }{0\% Strain} & \multicolumn{2}{ c }{6\% Strain} \\
\cline{3-6}
& & e & h & e & h \\
\hline
ZrS$_2$ & m$_{\Gamma M}$ & 2.543 & 0.608 & 4.175 & 0.720 \\
& m$_{M K}$ & 0.556 & 2.503 & 0.841 & 1.966 \\
\hline
ZrSe$_2$ & m$_{\Gamma M}$ & 2.247 & 0.455 & 3.250 & 0.658 \\
& m$_{M K}$ & 0.399 & 0.236 & 0.556 & 1.857 \\
\hline
ZrTe$_2$ & m$_{\Gamma M}$ & 2.075 & 0.273 & 2.591 & 0.351 \\
& m$_{M K}$ & 0.350 & 0.176 & 0.443 & 1.742 \\
\hline
\hline
\end{tabular}}%
\end{table}

The scattering parameters were chosen to be $r = \frac{3}{2}$ and $p = 0$ since these values are a typical trait of deformation potential scattering by alloys of 2D materials as mentioned in section \ref{ert}.

\begin{figure}[h!]
\includegraphics[scale=0.35]{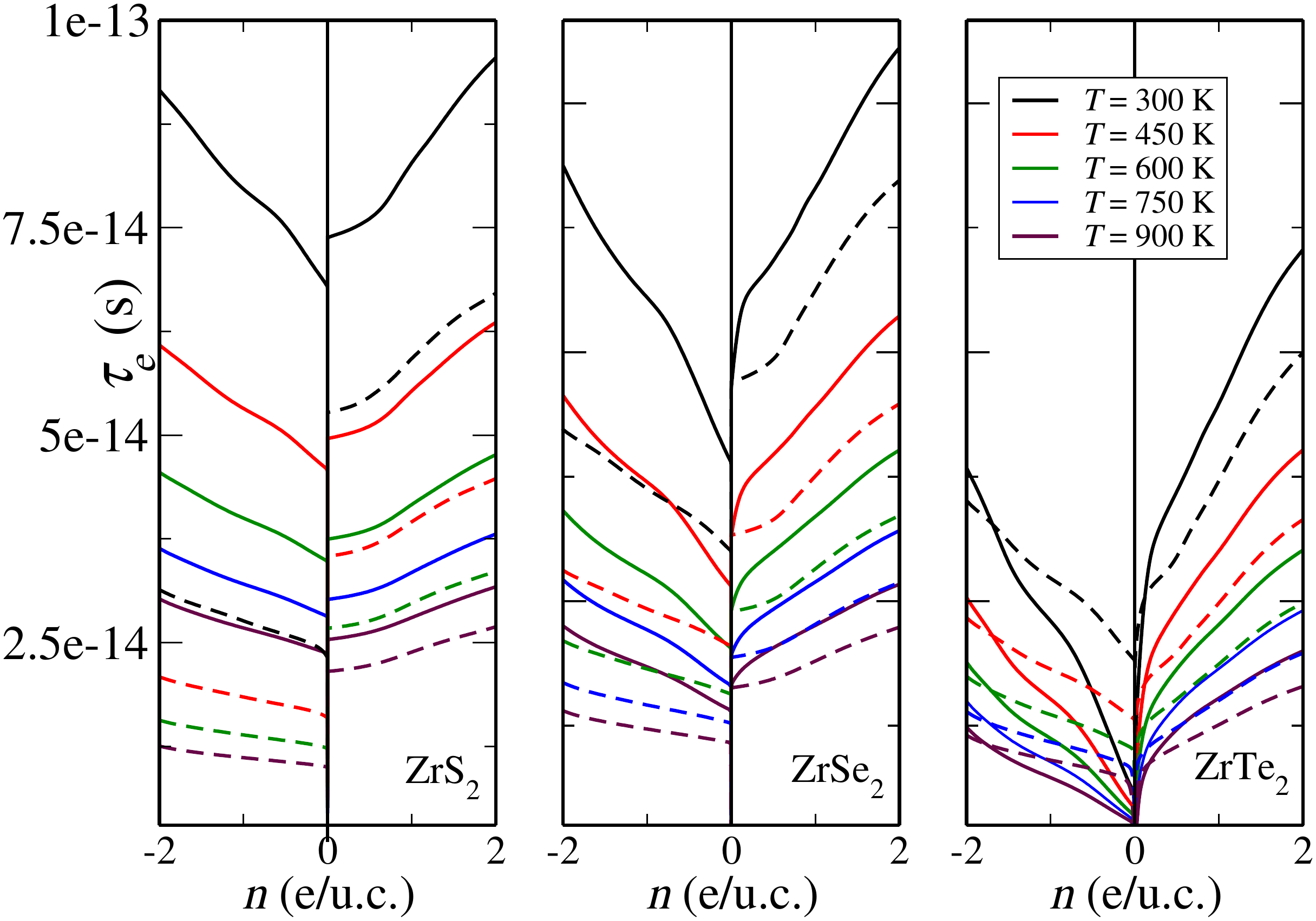}
\caption{The electron relaxation time as a functin of $n$-type and $p$-type charge carriers for unstrained and 6\% tensile strain ZrX$_2$. The straight (dotted) curves refer to that of the unstrained (6\% tensile strain) ZrX$_2$.}\label{tau_fig}
\end{figure}

In Fig. \ref{tau_fig}, we plot the relaxation time of ZrX$_2$ as a function of $n$-type and $p$-type charge carriers. There have been earlier reports estimating the relaxation time for ZrX$_2$ (X = S,Se) \cite{lv16,ding16} which are of the same order of magnitude with our calculations. However, those methods use a constant $\tau_e$ for a given temperature. Our calculated $\tau_e$, being a function of charge carrier (and hence chemical potential), would demonstrate characteristics observed experimentally. For example, the relaxation time for 2D graphene have been reported by Tan {\it et al.} \cite{tan07}. It can be seen that while the magnitude of the charge carriers increase, $\tau_e$ increases, in-line with our calculations. For semiconductors, the width of the bandgap is small enough for an electron from the valence band to enter the conduction band. Therefore, a larger concentration of electrons and holes would imply larger electrical conductivity and hence higher relaxation times. It can also be seen that for higher temperature the relaxation time decreases. This is due to the increased thermal velocity which increases the collision between electrons thus reducing $\tau_e$.
Under the impact of tensile strain, the relaxation time decreases for ZrX$_2$. This characteristic can be understood due to the increased effective mass as seen from table \ref{tab}.

\subsection{Lattice thermal Boltzmann Transport}
\begin{figure}[h!]
\centering
\includegraphics[scale=0.3]{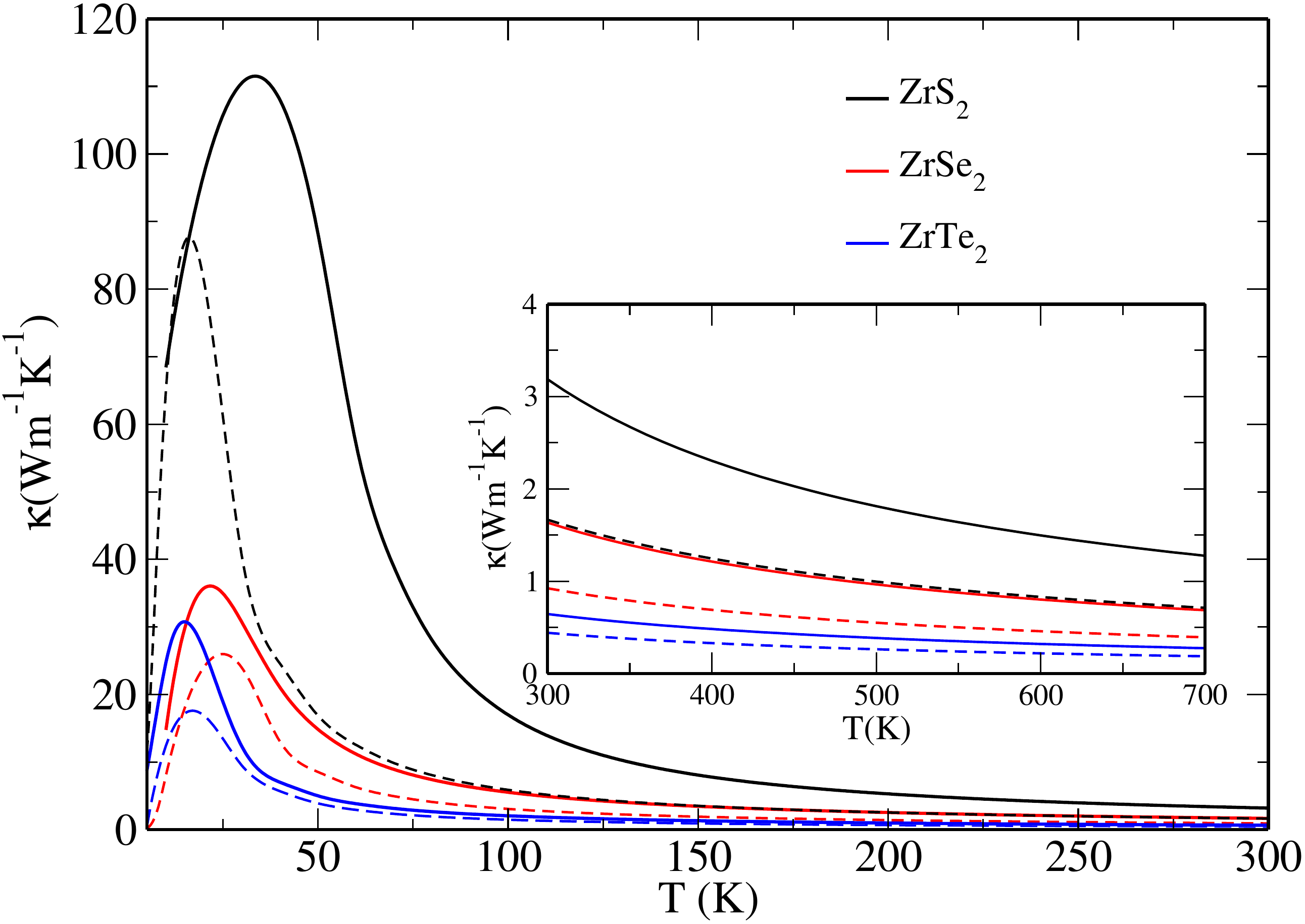}
\caption{Lattice thermal conductivity of unstrained (solid lines) and 6\% tensile strain (dashed lines) ZrX$_2$ as a function of temperature. }\label{kl-fig}
\end{figure}

The phonon dispersion plays a vital role in investigating the lattice thermal conductivity. Very recently, we calculated the phonon dispersion of unstrained and 6\% biaxial tensile ZrX$_2$ and found no imaginary frequencies in either of the systems thus confirming its thermal stability \cite{srs19}.
The temperature dependence of the lattice thermal conductivity ($\kappa_L$) of unstrained and 6\% tensile strain ZrX$_2$ is shown in Fig. \ref{kl-fig}.
For clarity, in the inset of Fig. \ref{kl-fig}, we plot $\kappa_L$ in the $T$=300 - 700 K temperature range.
It is observed that the variation of $\kappa_L$ is the maximum in the case of ZrS$_2$ followed by ZrSe$_2$ and ZrTe$_2$.
This behavior can be explained as follows.
The acoustic modes contribute almost entirely to $\kappa_L$. As we move from S to Se to Te, the maximum acoustic frequencies keep reducing. This results in lower group velocities.
Therefore, from Eq.\ref{kl}, having reduced $\omega$ and $v$ would result into lower $\kappa_L$.
By definition of the dynamical matrix, it can be seen that the phonon frequencies are proportional to the reciprocal of the atomic masses in the unit cell. It therefore follows that within a particular group, $\kappa_L$ would reduce if the constituent atoms are positioned comparatively lower in the periodic table.
For example, MoSe$_2$ \cite{kumar15} has a lower $\kappa_L$ in comparison to MoS$_2$ \cite{li15}. Similarly, very recently, Mobaraki {\it et al.} \cite {mobaraki18} showed that WSe$_2$ demonstrates a lower $\kappa_L$ relative to WS$_2$.
Moreover, ZrTe$_2$ would have an even smaller $\kappa_L$ due to the fact that the optical modes and
acoustic modes are coupled \cite{srs19} unlike in the case of ZrS$_2$, MoS$_2$, WS$_2$. Even the optical modes and acoustic modes are separate by a tiny frequency gap in ZrSe$_2$. In the small frequency range where the optical modes couple with the acoustic mode, the scattering rates in phonons would therefore increase thus resulting in an even lower $\kappa_L$.

\begin{figure}[h!]
\centering
\includegraphics[scale=0.3]{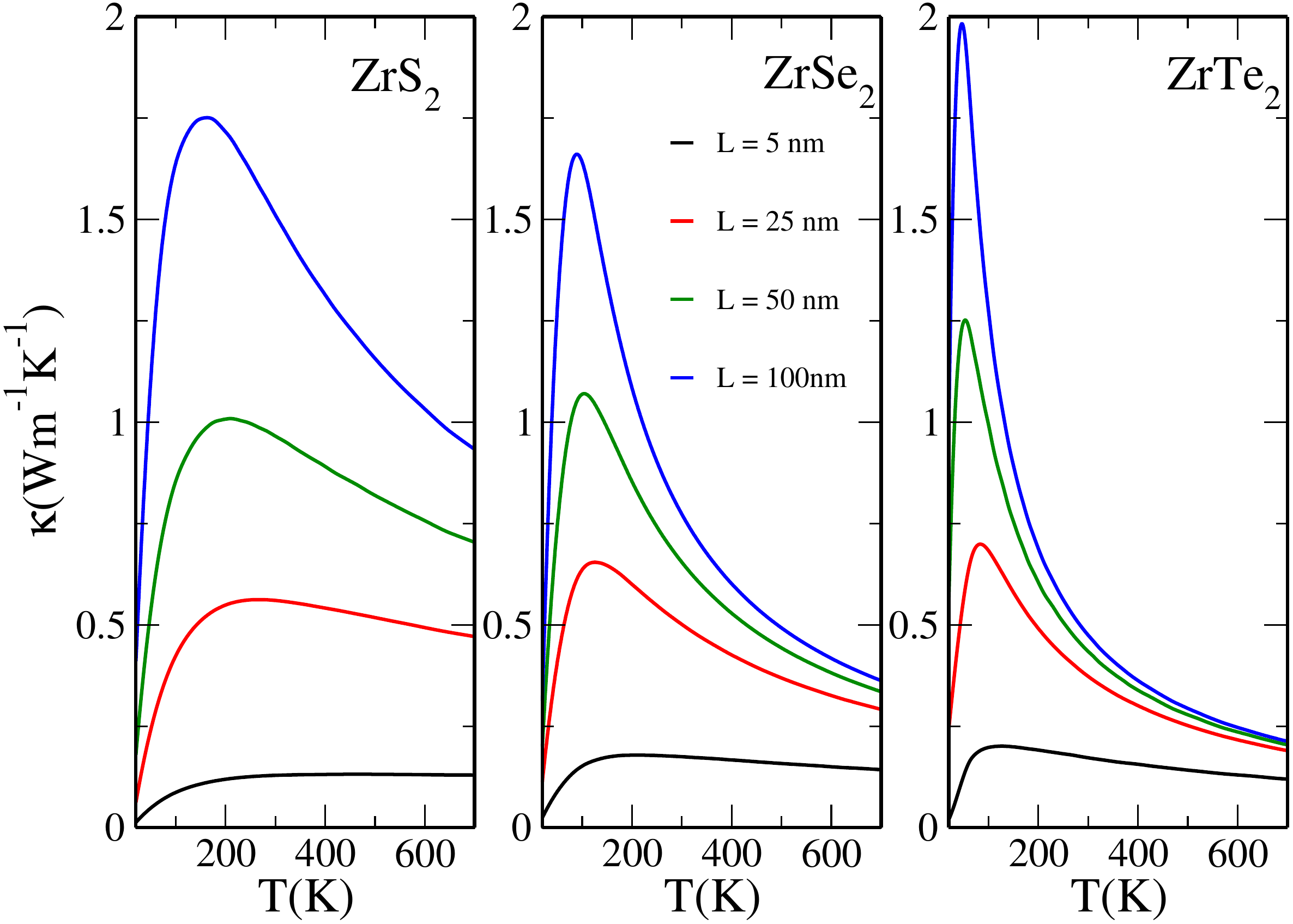}
\caption{Lattice thermal conductivity of ZrX$_2$ at constant lengths, 5 nm, 25nm, 50nm and 100nm. }\label{kl-length}
\end{figure}

Fig. \ref{kl-length}, shows the temperature dependence of the total thermal conductivity for different sample lengths, $L$= 5nm, 25nm, 50nm and 100nm. In all cases it is observed that, for the smallest length, $L$ = 5 nm, $\kappa_L$ initially increases but then at higher temperatures, $T > 100$ K, has a slight or no dependencies with temperature ($T$). The $T$ dependences increase as the sample length of the system increases. Around $L=$100nm, the $\frac{1}{T}$ dependence starts to form.
This characteristic can be understood by recalling the scattering rates that contribute to the total phonon relaxation time.
The boundary scattering $\frac{1}{\tau_B}$ is inversely proportional to length and has no temperature dependence while the anharmonic scattering rates have a $\frac{1}{T}$ dependence. Therefore by Matthiessen’s rule \cite{ashcroft}, the total phonon relaxation time is almost entirely dominated by the $\tau_B$ for extremely small sample lengths. With increase in the sample length, the contribution from the boundary scattering therefore decreases. The dominant scattering is now the anharmonic scattering rates and hence the $\frac{1}{T}$ dependencies of temperature start to form.

\begin{figure}[h!]
\centering
\includegraphics[scale=0.34]{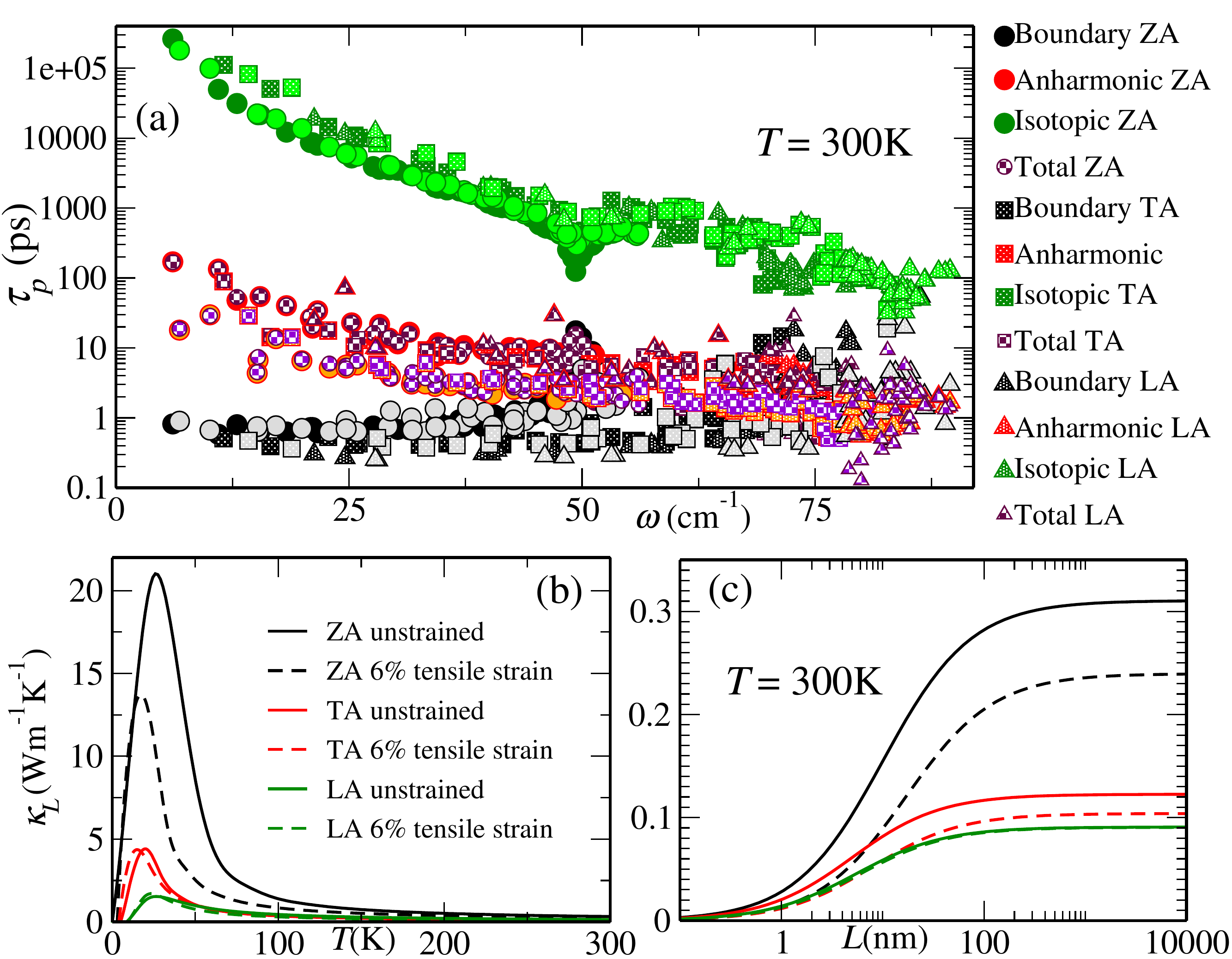}
\caption{(a): The phonon lifetime contributions, {\it i.e.}, boundary scattering(per unit length), anharmonic scattering and isotopic scattering, from each of the acoustic modes to the total thermal conductivity for ZrTe$_2$. Legends for the unstrained case are shown to the right of (a) whereas legends with the similar symbols but of different colours refer to the scattering rates for ZrTe$_2$ with 6\% strain.
(b and c): Acoustic mode contribution to the total lattice thermal conductivity (b) as a function of {\color{black}temperature} and (c) as a function of sample length at room temperature. The out-of-plane acoustic modes are denoted with black curves while the in-plane acoustic modes are denoted by red for the transverse acoustic (TA) modes and green for the longitudinal acoustic (LA) modes. Solid curves refer to unstrained system while the dotted curves correspond to the 6\% strain strained system.}\label{mode-kappa}
\end{figure}

In all of our calculations, the biaxial strain reduces the lattice thermal conductivity.
To have a better understanding of the influence of strain in $\kappa_L$, we inspect the mode dependence of ZrTe$_2$.
In Fig. \ref{mode-kappa} (a) we plot the phonon lifetimes contributed from boundary scattering (per mean free path), isotopic scattering and anharmonic scattering of each acoustic mode for unstrained and 6\% tensile strain ZrTe$_2$.
The phonon lifetimes are comparatively larger for smaller wavelengths indicating that heat is being transferred predominantly by acoustic phonons. Additionally, at smaller wavelengths, the phonon dispersions do not couple with each other and therefore the phonon-phonon scattering rates are low.
Our calculations clearly show that the dominant scattering is due to the anharmonic scattering rates.
Moreover, except for the anharmonic scattering rates from the ZA modes, the mode dependence calculation of phonon lifetimes show very little dependency with strain implying that tensile strain decreases the phonon lifetime of the ZA mode. 
The decrement of the anharmonic scattering occurring at lower frequencies can be inferred as follows.
Tensile strain on the ZA mode of ZrTe$_2$ transforms the quadratic behavior at lower frequencies to a more linear phonon dispersion thus decreasing the phonon density of states \citep{RDSM18} and therefore decreasing the phonon lifetimes of the ZA mode.
Our calculation of scattering rates could be verified using experimental techniques such as inelastic neutron or X-ray scattering\cite{pang13}

In Fig. \ref{mode-kappa} (b) we plot the acoustic mode contributions to $\kappa_L$.
The optical modes contribute negligibly to $\kappa_L$ since the phonon scattering rates are large and the group velocities are small in comparison with the acoustic modes.
The out-of-plane acoustic flexural modes (ZA) are shown in black. The transverse (TA) and longitudinal acoustic (LA) mode are shown in red and green, respectively. The solid and dotted lines refer to that of unstrained and 6\% biaxial tensile strain.

Our calculations demonstrates that the ZA modes contribute {\color{black}maximally} to $\kappa_L$ with little effects from TA and LA modes. The majority of ZA modes contributing to $\kappa_L$ is in line with the mode contributions seen in MoS$_2$ \cite{li15}.
The rotational symmetry of the out-of-plane mode which causes the quadratic nature of the ZA modes near the high symmetric $\Gamma$ point corresponds to the large density of flexural phonons.
An applied strain softens the in-plane (TA and LA) modes while it stiffens the out-of-plane modes.
A biaxial tensile strain gradually changes the {\color{black}quadratic} nature of the ZA mode to a linear nature thus implying a break in the rotational symmetry and resulting in a decrease in contribution from the ZA mode.
For a sufficiently large strain, the ZA modes becomes linear and will contribute less in comparison to the LA and TA mode. Such a trend has been seen in 2H-MoTe$_2$ \cite{shfique17}.
A tensile strain would always decrease the thermal conductivity if the system does not have reflection symmetry \cite{shfique17}.

In Fig. \ref{mode-kappa} (c), we plot the mode contribution to $\kappa_L$ as a function of length. The maximum increase in $\kappa_L$ is seen at lower lengths. As mentioned earlier, since $\tau_B$ dominates for small mean free paths, $\kappa_L$ increases linearly at smaller sample lengths. For larger lengths, the anharmonic scattering rates, which have no length dependence, dominate. This, therefore, results in a constant $\kappa_L$.
It is precisely because of the behaviour of $\kappa_L$ as a function of $L$ that nanostructuring and doping result in decreasing $\kappa_L$. Nanostructuring and doping further decrease the phonon mean free path and therefore must be operated at lengths where the boundary scattering dominate the total scattering rates.
It can also be seen from Fig. \ref{mode-kappa} (b and c) that summing the contribution from LA and TA would contribute almost the same amount to $\kappa_L$ as that of those from ZA modes. A strain larger than 6\% would therefore imply that the dominant contribution to $\kappa_L$ at room temperature would be from in-plane modes and not from the out-of-plane modes.

\subsection{Figure of Merit}
\begin{figure}[h!]
\includegraphics[scale=0.35]{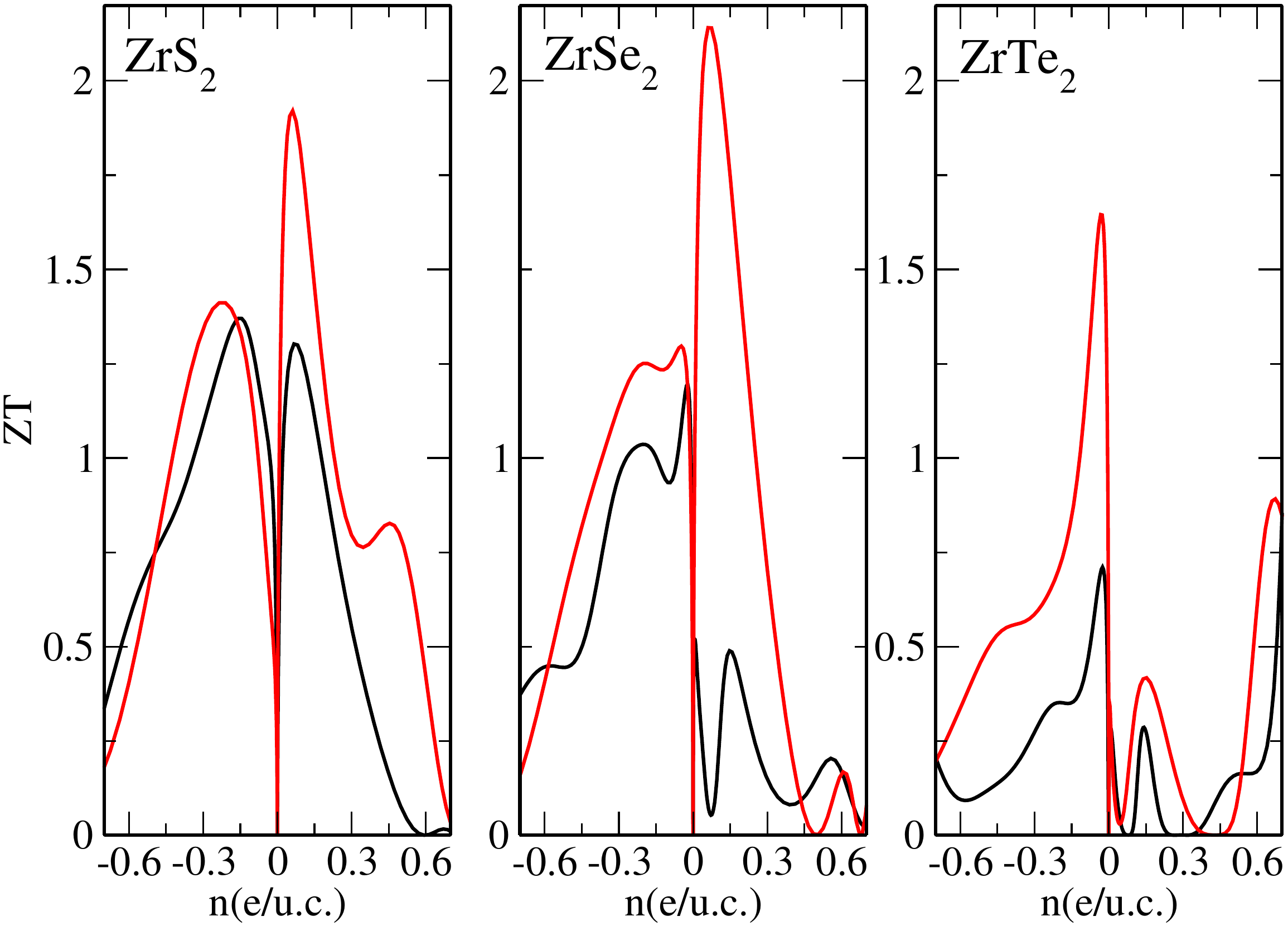}
\caption{The figure of merit of unstrained and 6\% tensile ZrX$_2$. The black curves refer to that of the unstrained systems while the red curves refer to that of the tensile strain. }\label{zt_strain}
\end{figure}

Having calculated all the thermoelectric parameters required for the conversion efficiency of ZrX$_2$,  we can now compute the figure of merit, $ZT = \frac{\tau_e (\frac{\sigma}{\tau_e}) S^2 T}{\kappa}$.
In Fig. \ref{zt_strain}, we plot the room temperature Figure of merit of unstrained and 6\% tensile strain ZrX$_2$ at the thermodynamic limit. Our calculations clearly show that $ZT$ of group {\rm IV} TMDs are good candidates for 2D thermoelectric materials. As expected, we find a large enhancement in $ZT$ when the system is under a biaxial tensile strain. The enhancement of $ZT$ is due to the simultaneous increase in powerfactor and decrease in the lattice thermal conductivity.
An interesting observation in our calculations revel that, for ZrS$_2$ and ZrSe$_2$, under tensile strain, the $p$-type doping results in a superior thermoelectric performance while the reverse is observed for $n$-type doping. This is in-line with another ZrS$_2$ report \cite{lv16}.
For the case in ZrTe$_2$, $n$-type doping would result in an enhanced $ZT$ with and without an impact in tensile strain.
The extent in the enhancement of $ZT$ in ZrS$_2$ in our calculation is smaller than that seen in a recent previous report \cite{lv16} since the effective mass calculated by us is larger thus reducing the relaxation times and hence the powerfactor.

\section{Conclusion}
To summarize, using first-principle methods, we have calculated all the thermoelectric parameters required to obtain the figure of merit of ZrX$_2$. Parameters related to electrons, such as, electrical conductivity, Seebeck coefficients and electron relaxation time were calculated using the electron Boltzmann equations on band electrons derived from density functional theory. The parameter related to phonons, {\it i.e.}, the lattice thermal conductivity was calculated using the phonon BTE from the phonon dispersion and inter atomic force constant extracted from density functional perturbation theory. Our calculations demonstrate that tensile strain in ZrX$_2$ would enhance the figure of merit by simultaneously increasing the powerfactor and reducing the lattice thermal conductivity.

Powerfactor calculations of unstrained ZrX$_2$ suggest $n$-type doped system yield better thermoelectric performance than their $p$-type counterpart.
Moreover, higher temperature yields higher powerfactor with the maximum corresponding to larger charge carriers due to excited electrons.
Under the influence of tensile strain, the maximum powerfactor as a function of charge carriers (or chemical potential) is found to be when the system has the maximum band gap.
We interpret these results using the Mott's formula which shows that the Seebeck coefficient is proportional to the inverse of electrical conductivity and the derivative of the electrical conductivity.
Since the shape of the bands do not change much, the Seebeck coefficient is highly dependent on the inverse electrical conductivity. Therefore, larger the band gap would imply lower conductivity and hence higher Seebeck coefficients and powerfactor.

Room temperature lattice thermal conductivity calculations show that the largest $\kappa_L$ is for ZrS$_2$ then ZrSe$_2$ and the lowest for ZrTe$_2$ since the highest acoustic frequencies and velocities keep reducing as we move from ZrS$_2$ to ZrSe$_2$ to ZrTe$_2$. Length dependent calculations show that there is little or no temperature dependence for $\kappa_L$ at small sample lengths ($L = 5$nm) while it displays a  $T^{-1}$ temperature dependence at larger sample lengths ($L = 100$nm). This is due to the boundary scattering rates which dominate at small sample lengths while the anharmonic scattering rates dominate at larger sample lengths.

Mode dependent calculations were carried out to understand the reduction in the lattice thermal conductivity due to tensile strain.
The in-plane longitudinal and transverse acoustic modes were not affected to a large extent due to tensile strain while the out of plane modes were greatly reduced. The calculations on phonon lifetimes also demonstrate that anharmonic scattering rates of ZA modes were affected the most due to tensile strain thus lowering the out of plane ZA phonon lifetime. Our mode dependent calculations also suggest that for strains larger than 6\%, the dominant modes contributing to the lattice thermal conductivity would be due the in-plane acoustic modes due to the stiffening of the ZA modes.

The electrical relaxation time was calculated from the effective mass. Larger the magnitude of the charge carrier masses, larger is the relaxation time, consistent with experimental observations. As for the dependence with temperature, the electrical relaxation time reduces with temperature because of its thermal velocity.

Finally, merging all the thermoelectric parameters together, we obtained the figure of merit of unstrained and 6\% tensile strain ZrX$_2$. For (un)strained ZrS$_2$ and ZrSe$_2$, $(n)p$-type doping would yield superior $ZT$
while for ZrTe$_2$, $n$-type doping would yield better conversion efficiency.
{\color{black}Our extensive study should provide useful information on the calculation of thermoelectric figure of merit from atomic positions alone since our calculations are free of any fitting parameters.}
Since strain can be induced experimentally in 2D materials \cite{gao07,park08,kim09}, and mode contribution to the lattice thermal conductivity can be measured \cite{minnich11,regner13,johnson13}, our calculations should motivate similar experimental studies in 2D ZrX$_2$ monolayers.

\section{Acknowledgments}
The calculations were performed in the High Performance Cluster platform at S.N. Bose National Centre (SNBNCBS), funded by the Department of Science and Technology. We are thankful to Peter Kratzer and Jesus Carrete for helpful correspondance. SA would like to express gratitude to King Khalid university, Abha, Saudi Arabia, for providing administrative and technical support and acknowledges support of the Visitors Programme (EVLP) during a visit to SNBNCBS.

\bibliographystyle{ieeetr}
\bibliography{zrx2-jap}

\end{document}